\documentclass[aps,prb,twocolumn,showpacs]{revtex4}
\usepackage{bm}
\usepackage{graphicx}
\usepackage{epsfig}
\begin{document}

\title{Electron scattering in quantum wells subjected to an in-plane
magnetic field}
\author{S.\,A.\,Tarasenko}
\affiliation{A.F.~Ioffe Physico-Technical Institute, Russian
Academy of Sciences, 194021 St.~Petersburg, Russia}


% 73.50.-h Electronic transport phenomena in thin films
% 73.50.Bk General theory, scattering mechanisms
% 73.50.Jt Galvanomagnetic and other magnetotransport effects
% (including thermomagnetic effects)

% 73.63.-b Electronic transport in nanoscale materials and
% structures (see also 73.23.-b Electronic transport in mesoscopic
% systems)
% 73.63.Hs Quantum wells

\begin{abstract}
It is shown that the electron scattering by static defects,
acoustic or optical phonons in quantum wells subjected to an
in-plane magnetic field is asymmetric. The probability of
scattering contains terms which are proportional to both the
electron wave vector and the magnetic field components. The terms
under study are caused by the lack of an inversion center in
quantum wells due to structure or bulk inversion asymmetry
although they are of pure diamagnetic origin. Such a magnetic
field induced asymmetry of scattering can be responsible for a
number of phenomena. In particular, the asymmetry of inelastic
electron-phonon interaction leads to an electric current flow if
only the electron gas is driven out of thermal equilibrium with
the crystal lattice.
\end{abstract}

\maketitle

\section{Introduction}

The processes of scattering of charge carriers by static defects,
acoustic and optical phonons, plasmons, and other quasiparticles
play an important role in solid states. They govern various
transport and optical properties of both bulk materials and
low-dimensional structures and determine, to a great extent, the
performance of semiconductor devices. Signification information on
the electron scattering in semiconductors is obtained by
magnetotransport and magnetooptical measurements. In quantum well
(QW) structures, the external magnetic field can be applied in the
interface plane or perpendicular to the plane. The effect of the
in-plane magnetic field on electron states in QWs is not so
pronounced as that of the perpendicular field because, due to the
strong quantum confinement, the in-plane field does not form the
Landau levels. However, it has been established that the in-plane
magnetic field leads not only to spin splitting of electron states
(the Zeeman effect) but also affects the orbital motion of free
carriers.~\cite{Ando82} In asymmetrical QW structures, the
in-plane field induces a diamagnetic shift of each electron
subband in $\bm{k}$-space,~\cite{Ando82} which becomes important
in tunneling between coupled QWs (see, e.g.,
Refs.~[\onlinecite{Demmerle91,Eisenstein91,Lin02,Raichev04,Vieira07}])
and direct optical
transitions,~\cite{Gorbatsevich93,Aleshchenko93,Kulik00,Ashkinadze05,Diehl07}
and can also modify the electron-phonon
interaction.~\cite{Kibis98,Kibis99}

Here we address the effect of an external magnetic field applied
in the interface plane on the electron scattering in quantum
wells. We show that the magnetic field leads to an in-plane
asymmetry of the electron scattering by static defects or phonons,
so that the scattering probability contains additional terms which
are proportional to both the electron wave vector and the magnetic
field components. Taking into account this contribution one can
write for the scattering rate
\begin{equation}\label{W_kk}
W_{\bm{k}'\bm{k}}=W_0 + \sum_{\alpha\beta} w_{\alpha\beta}
B_{\alpha} (k_{\beta} + k_{\beta}') \:,
\end{equation}
where $\bm{k}$ and $\bm{k}'$ are the initial and scattered wave
vectors, respectively, and $\bm{B}$ is the magnetic field. The
scattering rate at zero field $W_0$ and the pseudo-tensor
components $w_{\alpha\beta}$ are determined by the details of
scattering processes and, owing to the time inversion symmetry,
are even functions of the electron wave vector. Note, that in fact
the asymmetric part of the scattering rate can contain all terms
odd in the wave vector including cubic terms. Although the
scattering asymmetry does not modify the energy spectrum, it can
affect kinetics of the carriers.

Phenomenologically, asymmetric terms in the scattering rate given
by Eq.~(\ref{W_kk}) are caused by the lack of a spatial inversion
center in quantum wells due to structure and/or bulk inversion
asymmetry. This follows from the symmetry analysis which does not
require knowledge of microscopical mechanisms of the scattering.
Indeed, the point groups of noncentrosymmetrical QWs make no
difference between certain components of the axial vector $\bm{B}$
and the polar vectors $\bm{k}$ and $\bm{k}'$ allowing for the
coupling $B_{\alpha}(k_{\beta}+k'_{\beta})$. The sign "$+$" in
parenthesis is in accordance with the time inversion symmetry
imposing the condition
$W_{\bm{k}'\,\bm{k}}(\bm{B})=W_{-\bm{k}\:-\bm{k}'}(-\bm{B})$. We
note that, from the point of view of the symmetry analysis, the
terms under study are similar to spin-orbit induced contributions
to the matrix element of scattering
$\propto\sigma_{\alpha}(k_{\beta}+k'_{\beta})$,~\cite{Tarasenko06}
where the Pauli spin matrices $\sigma_{\alpha}$ are replaced by
the magnetic field components $B_{\alpha}$. However,
odd-in-$\bm{k}$ terms in Eq.~(\ref{W_kk}) are of pure diamagnetic
origin being not related to the spin of carriers.

To be specific, we consider below (001)-oriented QWs grown from
zinc-blende-type compounds. In such structures, the pseudo-tensor
$w_{\alpha\beta}$ contains two linearly independent components and
the scattering rate~(\ref{W_kk}) assumes the form
\begin{eqnarray}\label{W_001}
W_{\mathbf{k}'\mathbf{k}}=W_0 &+& w_{\mathrm{SIA}} [B_x (k_y+k_y')
- B_y (k_x+k_x')] \\
&+& w_{\mathrm{BIA}} [B_x (k_x+k_x') - B_y (k_y+k_y')] \nonumber
\:,
\end{eqnarray}
where the coefficients $w_{\mathrm{SIA}}$ and $w_{\mathrm{BIA}}$
are caused by structure and bulk inversion asymmetry,
respectively, $x \parallel [100]$ and $y \parallel [010]$ are the
in-plane coordinates, and $z \parallel [001]$ is the QW normal.

Microscopically, the scattering asymmetry originates from the
Lorentz force which acts on moving charge carriers and modifies
their wave functions. The effect is most easily conceivable for
the terms in the scattering rate caused by structure inversion
asymmetry, i.e., terms proportional to $w_{SIA}$ in
Eq.~(\ref{W_001}). This case is illustrated in Fig.~1 for the
electron scattering by impurities. The structure inversion
asymmetry is modeled here by placing the $\delta$-layer of
impurities (dotted line) closer to the lower interface rather than
exactly in the QW center. The magnetic field $\bm{B}$ is applied
along the $y$ axis. Electrons with different velocities $v_x =
\hbar k_x/m^*$ ($m^*$ is the effective mass) move in the QW plane.
Due to the Lorentz force $\bm{F}_L = (e/c)[\bm{v}\times\bm{B}]$,
where $e$ is the electron charge and $c$ is the light velocity,
the magnetic field $\bm{B}\parallel y$ pushes electrons to the
lower or upper interface depending on the sign of their velocity
$v_x$. This leads to modification of the function of size
quantization of electrons along the QW normal $\varphi(z)$ which
becomes $v_x$-dependent, as shown in Fig.~1. Since the
$\delta$-layer of impurities is shifted from the QW center to the
lower interface, the wave function of electrons with the positive
velocity $v_x$ is better overlapped with the impurity potentials
than the function of electrons with the negative $v_x$. As a
result, the electrons with $v_x>0$ are scattered by impurities at
higher rate than the carriers with $v_x<0$ leading to the in-plane
asymmetry of the scattering. Since the Lorentz force is
proportional to both the magnetic field and the electron velocity,
the small corrections to the scattering rate are linear in
$\bm{k}$ and linear in $\bm{B}$. In the above model, we assumed
that the structure inversion asymmetry and, consequently,
$\bm{k}$-linear terms in the scattering rate are caused by
asymmetry of the doping profile with respect to the QW center.
Obviously, the same arguments are valid for structures where
nonequivalence of $z$ and $-z$ directions is achieved by asymmetry
of the QW confinement potential.

\begin{figure}[t]
\leavevmode \epsfxsize=0.99\linewidth
\centering{\epsfbox{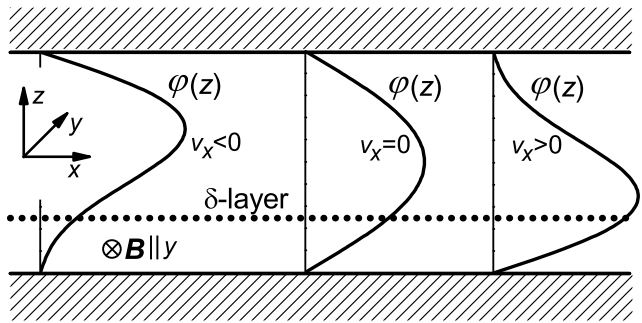}} \caption{Microscopic origin of the
in-plane asymmetry of electron scattering by $\delta$-layer of
impurities (dotted line) shifted with respect to the QW center.
Due to the Lorentz force, the magnetic field $\bm{B}\parallel y$
pushes electrons with the positive (negative) velocity $v_x$ to
the lower (upper) interface leading to increase (decrease) of the
scattering rate.} \vspace{-0.5cm}
\end{figure}

The term given by $w_{BIA}$ in Eq.~(\ref{W_001}) is related to the
lack of an inversion center of the QW host crystal with the
zinc-blende lattice. Microscopically, it is caused by modification
of the conduction-band wave function due to the magnetic field
induced inter-band mixing. We leave such contributions out of
scope of the present paper focusing below on the terms caused by
structure inversion asymmetry.

The rest of the paper is organized as follows. Section II is
devoted to microscopic calculations of asymmetric terms in the
scattering probability for different mechanisms of the electron
scattering including scattering by static defects, acoustic and
optical phonons. In Section III, we consider one of manifestations
of the scattering asymmetry: we show that asymmetry of the
electron-phonon interaction leads to the generation of an electric
current if only the electron gas is driven out of thermal
equilibrium with the crystal lattice by any means. We shall focus
on diamagnetic effects only and, therefore, neglect spin related
contributions for simplicity.

\section{Microscopic theory}

We consider linear in the magnetic field terms in the scattering
probability caused by structure inversion asymmetry. To calculate
such terms, it is sufficient to consider one-band model and assume
that the contribution to the effective Hamiltonian induced by the
in-plane magnetic field is given by
\begin{equation}\label{H_dia}
H_{\bm{B}} = \frac{e\hbar}{m^*c} (B_x k_y - B_y k_x) z \:,
\end{equation}
where $z$ is the coordinate operator.~\cite{Ando82} The
Hamiltonian~(\ref{H_dia}) corresponds to the magnetic field
$\bm{B}=\mathrm{rot} \bm{A}$ with the vector potential chosen in
the form $\bm{A} = (B_y z, -B_x z, 0)$.

The in-plane magnetic field intermixes electron states from
different quantum subbands at nonzero wave vector $\bm{k}$ leading
to dependence of the functions of size quantization on the wave
vector.~\cite{Falko02} To first order in the perturbation theory,
the electron wave function of the first level of size quantization
$e1$ has the form
\begin{equation}\label{psi}
\psi_{1\bm{k}}(\bm{r}) = \varphi_{1\bm{k}}(z)
\exp(i\bm{k}\cdot\bm{\rho}) \:,
\end{equation}
where $\bm{\rho}$ is the in-plane coordinate,
$\varphi_{1\bm{k}}(z)$ is the function of size quantization,
\begin{equation}\label{varphi}
\varphi_{1\bm{k}}(z) = \varphi_{1}(z) - (B_x k_y - B_y k_x)
\frac{e\hbar}{m^* c} \sum_{\nu \neq 1} \frac{z_{\nu
1}}{\varepsilon_{\nu 1}} \,\varphi_{\nu}(z) \:,
\end{equation}
$\nu$ is the subband index [in Eq.(\ref{varphi}) it runs over all
conduction subbands except the subband $e1$], $\varepsilon_{\nu 1}
= \varepsilon_{\nu}-\varepsilon_{1}$ is the energy separation
between the subbands, $\varepsilon_{\nu}$ is the bottom energy of
the subband $\nu$, $z_{\nu 1}=\int \varphi_{\nu}(z) z
\varphi_{1}(z) dz$ is the coordinate matrix element, and
$\varphi_{\nu}$ is the function of size quantization of the
subband $\nu$ at zero magnetic field.

Taking into account the dependence of the electron wave functions
on the magnetic field~(\ref{varphi}), one can derive that the
scattering rate $W_{\bm{k}'\bm{k}}$ has the form of
Eq.~(\ref{W_001}), with the ratio between $w_{SIA}$ and $W_0$
being
\begin{equation}\label{WSIA_W0}
w_{SIA} = - 2 W_0 \frac{e\hbar}{m^*c} \sum_{\nu \neq 1}
\frac{z_{\nu 1}} {\varepsilon_{\nu 1}} \xi_{\nu 1} \:,
\end{equation}
where $\xi_{\nu 1}$ are dimensionless parameters. The explicit
form of the scattering rate at zero magnetic field $W_0$ and the
parameters $\xi_{\nu 1}$ depend on the details of scattering and
are calculated below.

\subsection{Scattering by static defects}

At low temperatures, the electron scattering is dominated by
elastic processes from static defects such as impurities,
imperfections in the QW interfaces, etc. In the case of
short-range static defects, the matrix element of scattering has
the form
\begin{equation}\label{V_imp}
V_{\bm{k}'\bm{k}} = V_0 \sum_{j} \int
\psi_{1\bm{k}'}^{*}(\bm{r})\delta(\bm{r}-\bm{r}_j)
\psi_{1\bm{k}}^{}(\bm{r}) d\bm{r} \:,
\end{equation}
where $V_0$ is a parameter characterizing the impurity strength,
$\bm{r}_{j}$ is the impurity position, and the index $j$
enumerates impurities contributing to the scattering.

The transition rate between electron states with the wave vectors
$\bm k$ and $\bm k'$ are defined, in the Born approximation, by
the standard equation
\begin{equation}
W_{\bm{k}'\bm{k}} = \frac{2\pi}{\hbar} |V_{\bm{k}'\bm{k}}|^2
\delta(\varepsilon_{\bm{k}'}-\varepsilon_{\bm{k}}) \:,
\end{equation}
where $\varepsilon_{\bm{k}}=\hbar^2k^2/(2m^*)$ is the electron
kinetic energy. We note that the small diamagnetic correction to
the kinetic energy, which is given by the diagonal matrix element
of the Hamiltonian~(\ref{H_dia}), is neglected in our analysis
since it leads to no essential contribution to the scattering
asymmetry. In fact, to first order in the magnetic field, it
results only in a displacement of the subband spectrum in
$\bm{k}$-space. This shift does not disturb the symmetric
distribution of carriers within the subband and can be excluded by
a proper choice of the coordinate origin.

Squaring the matrix element of scattering~(\ref{V_imp}) and
averaging it over the positions of impurities, one derives
\begin{equation}\label{W0_imp}
W_0 = \frac{2\pi}{\hbar} |V_0|^2 N_{d} \,
\delta(\varepsilon_{\bm{k}'}-\varepsilon_{\bm{k}}) \times
\int_{-\infty}^{\infty} \varphi_{1}^4(z)u(z)dz \:,
\end{equation}
\begin{equation}\label{xi_imp}
\xi_{\nu 1} = \frac{\int_{-\infty}^{\infty}
\varphi_{1}^3(z)\varphi_{\nu}(z)u(z)dz}{\int_{-\infty}^{\infty}
\varphi_{1}^4(z)u(z)dz} \:,
\end{equation}
where $N_{d}$ is the sheet density of impurities and $u(z)$ is
their distribution function along the growth direction, $\int u(z)
dz=1$.

\subsection{Scattering by acoustic phonons}

At finite temperatures, electron-phonon interaction can
predominate over the electron collisions with static defects. In
the case of scattering from bulk acoustic phonons, the squared
matrix element of the scattering assisted by emission or
absorption of a longitudinal phonon has the form~\cite{Ivchenko05}
\begin{equation}\label{V_ac}
|V_{\bm{k}'\bm{k}}^{\pm}(\bm{q})|^2 = \Xi_c^2 \frac{\hbar q^2
N_{\bm{q}}^{\pm}}{2\rho_c \,\Omega_{\bm{q}}} \left| \int
\psi_{1\bm{k}'}^{*}(\bm{r}) \mathrm{e}^{\mp i \bm{q}\cdot\bm{r}}
\psi_{1\bm{k}}(\bm{r})
 d\bm{r} \right|^2 \:.
\end{equation}
Here the upper and lower signs correspond to the phonon emission
and absorption, respectively, $\Xi_c$ is the conduction-band
deformation-potential constant, $\rho_c$ is the crystal density,
$N_{\bm{q}}^{-}=N_{\bm{q}}$, $N_{\bm{q}}^{+}=N_{\bm{q}}+1$,
$N_{\bm{q}}=1/[\exp{(\hbar\Omega_{\bm{q}}/k_B T_0)}-1]$ is the
phonon occupation number, $\Omega_{\bm{q}} \approx s_L q$ is the
frequency of the longitudinal acoustic wave, $k_B$ is the
Boltzmann constant, $T_0$ is the lattice temperature, $s_L$ is the
sound velocity,  $q=|\bm{q}|$, and $\bm{q}$ is the
three-dimensional wave vector of the phonon involved,
$\bm{q}=\pm(\bm{k}-\bm{k}',q_z)$.

The rate of the electron scattering assisted by emission or
absorption of photons is given by
\begin{equation}
W_{\bm{k}'\bm{k}}^{\pm}=\frac{2\pi}{\hbar}\sum_{\bm{q}}
|V_{\bm{k}'\bm{k}}^{\pm}(\bm{q})|^2 \delta(\varepsilon_{\bm{k}'}
-\varepsilon_{\bm{k}}\pm \hbar\Omega_{\bm{q}}) \:.
\end{equation}
We assume that electrons populate the bottom of the ground subband
$e1$. Then, one derives
\begin{equation}\label{W0_ac}
W_0^{\pm} = \frac{\Xi_c^2}{2 s_L \rho_c} \int_{-\infty}^{\infty}
|Q_{11}|^2 N_{\bm{q}}^{\pm} \,\delta(\varepsilon_{\bm{k}'}
-\varepsilon_{\bm{k}}\pm \hbar\Omega_{\bm{q}}) \,q\, dq_z \:,
\end{equation}
\begin{equation}\label{xi_ac}
\xi_{\nu 1}^{\pm} = \frac{\int_{-\infty}^{\infty}
\mathrm{Re}[Q_{11}Q_{\nu 1}^*] N_{\bm{q}}^{\pm}
\,\delta(\varepsilon_{\bm{k}'} -\varepsilon_{\bm{k}}\pm
\hbar\Omega_{\bm{q}}) \,q\, dq_z}{\int_{-\infty}^{\infty}
|Q_{11}|^2 N_{\bm{q}}^{\pm} \,\delta(\varepsilon_{\bm{k}'}
-\varepsilon_{\bm{k}}\pm \hbar\Omega_{\bm{q}}) \,q\, dq_z} \:,
\end{equation}
where $Q_{\nu 1}=\int \varphi_{\nu}(z) \varphi_{1}(z) \exp(i q_z
z) dz$.

The electron scattering by acoustic phonons is usually considered
as a quasi-elastic process because the energy of the phonon
involved $\hbar\Omega_{\bm{q}}$ is small as compared to the
electron kinetic energy $\varepsilon_{\bm{k}}$. If one is
interested in the momentum scattering only neglecting energy
transfer between the electron and phonon systems, one can omit the
term $\hbar\Omega_{\bm{q}}$ in the $\delta$-functions
in~Eqs.~(\ref{W0_ac})~and~(\ref{xi_ac}). Under this approximation
and provided $N_q^{\pm}\approx k_B T_0 /(\hbar\Omega_{\bm{q}})\gg
1$, the transition rate $W_0^{\pm}$ and the parameters $\xi_{\nu
1}^{\pm}$ assume the form
\[
W_0^{\pm} \approx \frac{\pi}{\hbar} \, \frac{k_B T_0}{s_L^2
\rho_c} \, \Xi_c^2 \,
\delta(\varepsilon_{\bm{k}'}-\varepsilon_{\bm{k}}) \times
\int_{-\infty}^{\infty} \varphi_{1}^4(z)dz \:,
\]
\[
\xi_{\nu 1}^{\pm} \approx \frac{\int_{-\infty}^{\infty}
\varphi_{1}^3(z)\varphi_{\nu}(z)dz}{\int_{-\infty}^{\infty}
\varphi_{1}^4(z)dz} \:,
\]
which is similar to Eqs.~(\ref{W0_imp}),~(\ref{xi_imp}) with
$u(z)=\mathrm{const}$. It is reasonable that acoustic phonons in
the deformation-potential model behave as an ensemble of
short-range scatterers uniformly distributed along the QW growth
direction.

\subsection{Scattering by optical phonons}

At even higher temperatures, the electron scattering is governed
by interaction with optical phonons. For the Fr\"{o}lich mechanism
of electron-phonon interaction,~\cite{Ivchenko05} the squared
matrix element of the scattering assisted by emission or
absorption of a longitudinal optical ($LO$) phonon has the form
\begin{equation}\label{V_opt}
|V_{\bm{k}'\bm{k}}^{\pm}(\bm{q})|^2 = \frac{2\pi e^2 \hbar
\Omega_{LO}N_{LO}^{\pm}}{\epsilon^* q^2} \left| \int
\psi_{1\bm{k}'}^{*}(\bm{r}) \mathrm{e}^{\mp i \bm{q}\cdot\bm{r}}
\psi_{1\bm{k}}(\bm{r}) d\bm{r} \right|^2 ,
\end{equation}
where $\Omega_{LO}$ is the phonon frequency, $1/\epsilon^* =
1/\epsilon_{\infty}-1/\epsilon_{0}$, $\epsilon_{0}$ and
$\epsilon_{\infty}$ are the dielectric constants at low and high
frequencies, respectively, $N_{LO}^{-}=N_{LO}$,
$N_{LO}^{+}=N_{LO}+1$, and $N_{LO}$ is the phonon occupation
number. We assume that electrons populate the bottom of the ground
subband only and the phonon energy $\hbar\Omega_{LO}$ is much
smaller than the energy separation between quantum subbands. In
this particular case, one obtains
\begin{equation}\label{W0_opt}
W_0^{\pm} = \frac{2 \pi^2 e^2}{\epsilon^*}
\frac{\Omega_{LO}}{|\bm{k}'-\bm{k}|} N_{LO}^{\pm} \,
\delta(\varepsilon_{\bm{k}'} -\varepsilon_{\bm{k}}\pm
\hbar\Omega_{LO}) \:,
\end{equation}
\begin{equation}\label{xi_opt}
\xi_{\nu 1} = - |\bm{k'}-\bm{k}| \int\int \varphi_1^2(z)
\varphi_1(z') \varphi_{\nu}(z') |z-z'| dzdz' \:.
\end{equation}

In accordance with general symmetry arguments, the asymmetric
terms in the scattering rate proportional to $w_{SIA}$ are related
to inversion asymmetry of the heterostructure and vanish for the
absolutely symmetrical quantum well. This follows also from
Eq.~(\ref{WSIA_W0}) together with
Eqs.~(\ref{xi_imp}),~(\ref{xi_ac}), or~(\ref{xi_opt}) which
demonstrate that the sign and magnitude of $w_{SIA}$ are
determined by the products $z_{\nu 1} \xi_{\nu 1}$. The products
are non-zero to the extent of asymmetry of the confinement
potential and/or the doping profile and vanish for the absolutely
symmetrical structure, where $u(z)$ is an even function and
$\varphi_{\nu}$ is either even or odd function with respect to the
QW center.

Following Eqs.~(\ref{W_001})~and~(\ref{WSIA_W0}) we can estimate
the ratio between the asymmetric and symmetric parts of the
scattering rate as $w_{SIA} B k/W_0 \propto (\hbar\omega_c
/\varepsilon_{21})k z_{21} \xi_{21}$, where $\omega_c=|e|B/(m^*
c)$. The estimate gives $w_{SIA} B k/W_0 \sim 10^{-3}$ for
GaAs-based QW structure with the asymmetry degree $\xi_{21}=0.1$,
$\varepsilon_{21}=100$~meV, and $kz_{21}=1$ in the magnetic field
$B=1$~T.

\section{Electric current caused by energy relaxation of carriers}

The heating of a two-dimensional electron gas subjected to an
in-plane magnetic field can lead to the generation of an electric
current. Such an effect has been observed in experiments where the
electron gas was driven out of thermal equilibrium with the
crystal lattice by a low-frequency electric
current~\cite{Pogosov00} or far infrared
radiation.~\cite{Belkov05} It has been shown that different
microscopic mechanisms including
diamagnetic~\cite{Kibis99,Pogosov00,Belkov05} as well as
paramagnetic
(spin-dependent)~\cite{Belkov05,Ivchenko83,Ganichev06} mechanisms
can be responsible for the effect. The magnetic field induced
asymmetry of the inelastic electron scattering by phonons given by
Eq.~(\ref{W_kk}) also gives rise to an electric current if the
electron gas temperature $T_e$ differs from the lattice
temperature $T_0$. This mechanism seems to be the dominant
diamagnetic contribution because the current originating from the
magnetic field induced modification of the energy spectrum
vanishes in $\bm{k}$-linear approximation and arises in high-order
approximations only.

The electric current in the relaxation time approach is given by
\begin{equation}\label{j_gen}
\bm{j} = 2 e \sum_{\bm{k}} \tau_p \, \bm{v} \dot{f}_{\bm{k}} \:,
\end{equation}
where $\tau_p$ is the moment relaxation time, $\dot{f}_{\bm{k}}$
stands for the generation function stemming from the electron
scattering by phonons, and the factor 2 in Eq.~(\ref{j_gen})
accounts for the spin degeneracy. The function $\dot{f}_{\bm{k}}$
has the form
\begin{equation}\label{f_dot}
\dot{f}_{\bm{k}} = \sum_{\bm{k}',\pm} [W_{\bm{k}\bm{k}'}^{\pm}
f_{\bm{k}'}(1-f_{\bm{k}}) - W_{\bm{k}'\bm{k}}^{\pm}
f_{\bm{k}}(1-f_{\bm{k}'})] \:,
\end{equation}
where $f_{\bm{k}}=1/\{\exp[(\varepsilon_{\bm{k}}-\mu)/(k_B
T_e)]+1\}$ is the distribution function of carriers, and $\mu$ is
the chemical potential. Combining Eqs.~(\ref{j_gen}) and
(\ref{f_dot}), one obtains for the current
\begin{equation}\label{j_gen2}
\bm{j} = 2 e \sum_{\bm{k}\bm{k}'} \tau_p (\bm{v}_{\bm{k}} -
\bm{v}_{\bm{k}'}) [W_{\bm{k}\bm{k}'}^{+} f_{\bm{k}'}(1-f_{\bm{k}})
- W_{\bm{k}'\bm{k}}^{-} f_{\bm{k}}(1-f_{\bm{k}'})] \:.
\end{equation}
In thermal equilibrium, when the electron and lattice temperatures
coincide, the expression in square brackets in Eq.~(\ref{j_gen2})
vanishes because the processes of phonon emission and absorption
compensate each other. In contrast, if $T_e \neq T_0$, the
absorption and emission rates become nonequal leading, due to
scattering asymmetry in $\bm{k}$-space, to the electric current.

We assume that the electron temperature differs slightly from the
lattice temperature, so that the inequality $|\Delta T|\hbar
\Omega_{\bm{q}} \ll k_B T_e T_0$ is fulfilled, where $\Delta T=T_e
- T_0$. Then, taking into account that the rate of electron
scattering assisted by emission or absorption of a phonon with the
wave vector $\bm{q}$ is proportional to $N_{\bm{q}}^{\pm}$ and
\[
\left. N_{\bm{q}}^+ f_{\bm{k}'}(1-f_{\bm{k}}) - N_{\bm{q}}^-
f_{\bm{k}}(1-f_{\bm{k}'})
\right|_{\varepsilon_{\bm{k}'}=\varepsilon_{\bm{k}}+\hbar\Omega_{\bm{q}}}
\approx
\]
\vspace{-0.5cm}
\[
N_{\bm{q}}^- f_{\bm{k}}(1-f_{\bm{k}'}) \frac{\Delta T}{T_e}
\frac{\hbar\Omega_{\bm{q}}}{k_B T_0} \:,
\]
one can derive the expression for the electric current valid in
the linear approximation in $\Delta T/T_e$.

Calculations show that the electric current caused by structure
inversion asymmetry flows in the direction perpendicular to the
applied magnetic field and in given by
\begin{equation}
j_x = j B_y /B \:, \;\; j_y = - j B_x /B \:,
\end{equation}
where $j$ is the electric current magnitude. We assume that the
temperature is sufficiently high and the carriers obey the
Boltzmann statistics. Then, for the scattering by acoustic
phonons, one derives
\begin{equation}\label{j_acoust}
j = \tau_p B N_e \frac{e^2 \Xi_c^2}{c \hbar \rho_c} \frac{\Delta
T}{T_e} \sum_{\nu \neq 1} \frac{z_{1\nu}}{\varepsilon_{\nu 1}}
\int \varphi_1(z) \varphi_{\nu}(z) \frac{d^2 \varphi_1^2(z)}{dz^2}
dz \:,
\end{equation}
where $N_e = 2 \sum_{\bm{k}} f_{\bm{k}}$ is the electron density.
In the case of the electron scattering by optical phonons, the
calculation yields
\begin{equation}\label{j_opt}
j = \tau_p B N_e \frac{2 \pi e^4}{c \,\epsilon^*}
\frac{\Omega_{LO}^3 N_{LO}}{k_B T_0} \frac{\Delta T}{T_e}
\sum_{\nu \neq 1} \frac{z_{1\nu}}{\varepsilon_{\nu 1}}
\end{equation}
\vspace{-0.7cm}
\[
\times \int\int \varphi_1^2(z) \varphi_1(z') \varphi_{\nu}(z')
|z-z'| dzdz'  \:.
\]
The estimation after Eq.~(\ref{j_acoust}) gives $j\sim0.1$~$\mu$A
in the magnetic field $B=1$~T for GaAs-based structures with the
momentum relaxation time $\tau_p=10^{-12}$~s, the carrier density
$N_e=10^{12}$~cm$^{-2}$, the relative temperature difference
$\Delta T/T_e=0.1$, and the QW asymmetry degree $\xi=0.1$.

In conclusion, we have shown that the magnetic field applied in
the quantum well plane has a diamagnetic influence on the
scattering of charge carriers. The magnetic field leads to odd in
the wave vector terms in the scattering rate resulting in the
in-plane asymmetry of the scattering. The scattering rate has been
calculated for the electron interaction with impurities as well as
acoustic and optical phonons.

\paragraph*{Acknowledgments.} This work was supported by the RFBR,
programs of the RAS, and the President Grant for young scientists.


\begin{thebibliography}{99}
\bibitem{Ando82} T.~Ando, A.B.~Fowler, and F.~Stern,
{\it Electronic properties of two-dimensional systems}, \rmp {\bf
54}, 437 (1982).

% Tunneling
\bibitem{Demmerle91} W.~Demmerle, J.~Smoliner, G.~Berthold, E.~Gornik,
G.~Weimann, and W.~Schlapp, {\it Tunneling spectroscopy in
barrier-separated two-dimensional electron-gas systems}, \prb {\bf
44}, 3090 (1991).
\bibitem{Eisenstein91} J.P.~Eisenstein, T.J.~Gramila, L.N.~Pfeiffer, and
K.W.~West, {\it Probing a two-dimensional Fermi surface by
tunneling}, \prb {\bf 44}, 6511 (1991).
\bibitem{Lin02} Y.~Lin, E.E.~Mendez, and A.G.~Abanov, {\it Tunneling
characteristics of an electron-hole trilayer in a parallel
magnetic field}, \prb {\bf 66}, 195311 (2002).
\bibitem{Raichev04} O.E.~Raichev and F.T.~Vasko, {\it Spin-polarized
tunneling current between independently contacted quantum wells},
\prb {\bf 70}, 075311 (2004).
\bibitem{Vieira07} G.S.~Vieira, W.H.M.~Feu, J.M.~Villas-B\^{o}as,
P.S.S.~Guimar\~{a}es, and N.~Studart, {\it Resonant tunneling
between thermal excited states tuned by a magnetic field}, \prb
{\bf 75}, 193406 (2007).

% Optical transitions
\bibitem{Gorbatsevich93} A.A.~Gorbatsevich, V.V.~Kapaev, and
Yu.V.~Kopaev, {\it Asymmetric nanostructures in a magnetic field},
Pis'ma Zh. Eksp. Teor. Fiz. {\bf 57}, 565 (1993) [JETP Lett. {\bf
57}, 580 (1993)].
\bibitem{Aleshchenko93} Yu.A.~Aleshchenko, I.D.~Voronova, S.P.~Grishechkina,
V.V.~Kapaev, Yu.V.~Kopaev, I.V.~Kucherenko, V.I.~Kadushkin, and
S.I.~Fomichev, {\it Magnetic-field-induced photovoltaic effect in
an asymmetric system of quantum wells}, Pis'ma Zh. Eksp. Teor.
Fiz. {\bf 58}, 377 (1993) [JETP Lett. {\bf 58}, 384 (1993)].
\bibitem{Kulik00} L.V.~Kulik, I.V.~Kukushkin, V.E.~Kirpichev,
K.~v.~Klitzing, and K.~Eberl, {\it Magnetic-field-induced
dispersion anisotropy of intersubband excitations in an
asymmetrical quasi-two-dimensional electron system}, \prb {\bf
61}, 1712 (2000).
\bibitem{Ashkinadze05} B.M.~Ashkinadze, E.~Linder, E.~Cohen, and
L.N.~Pfeiffer, {\it Effect of an in-plane magnetic field on the
photoluminescence spectrum of modulation-doped quantum wells and
heterojunctions}, \prb {\bf 71}, 045303 (2005).
\bibitem{Diehl07} H.~Diehl, V.A.~Shalygin, S.N.~Danilov, S.A.~Tarasenko,
V.V.~Bel'kov, D.~Schuh, W.~Wegscheider, W.~Prettl, S.D.~Ganichev,
{\it Magneto-gyrotropic photogalvanic effects due to inter-subband
absorption in quantum wells}, J. Phys.: Condens. Matter {\bf 19},
436232 (2007).

% Electron-phonon interaction
\bibitem{Kibis98} O.V.~Kibis, {\it Possible new quantum macroscopic effect
in low-dimensional structures: The appearance of an electromotive
force in a standing acoustic wave}, Physics Letters A {\bf 237},
292 (1998).
\bibitem{Kibis99} O.V.~Kibis, {\it Novel effects of electron-phonon interaction
in quasi-two-dimensional structures located in a magnetic field},
Zh. Eksp. Teor. Fiz. {\bf 115}, 959 (1999) [JETP {\bf 88}, 527
(1999)].

\bibitem{Tarasenko06} S.A.~Tarasenko, {\it Spin orientation of a
two-dimensional electron gas by a high-frequency electric field},
\prb {\bf 73}, 115317 (2006).
\bibitem{Falko02} V.I.~Fal'ko and T.~Jungwirth, {\it Orbital effect
of an in-plane magnetic field on quantum transport in chaotic
lateral dots}, \prb {\bf 65}, 081306 (2002).
\bibitem{Ivchenko05} E.L.~Ivchenko, Optical Spectroscopy of Semiconductor
Nanostructures (Harrow, UK: Alpha Science Int., 2005).

% Heating experiments
\bibitem{Pogosov00} A.G.~Pogosov, M. V.~Budantsev , O.V.~Kibis,
A.~Pouydebasque, D.K.~Maude, J.C.~Portal, {\it Thermomagnetic
effect in a two-dimensional electron gas with an asymmetric
quantizing potential}, \prb {\bf 61}, 15603 (2000).
\bibitem{Belkov05} V.V.~Bel'kov, S.D.~Ganichev,  E.L.~Ivchenko, S.A.~Tarasenko,
W.~Weber, S.~Giglberger,  M.~Olteanu, P.~Tranitz, S.N.~Danilov,
Petra~Schneider, W.~Wegscheider, D.~Weiss, and W.~Prettl, {\it
Magneto-gyrotropic photogalvanic effects in semiconductor quantum
wells}, J. Phys.: Condens. Matter {\bf 17}, 3405 (2005).

\bibitem{Ivchenko83} E.L.~Ivchenko and G.E.~Pikus, {\it Optical orientation
of free carriers spins and photogalvanic effects in gyrotropic
crystals}, Izv. Akad. Nauk SSSR, Ser. Fiz. {\bf 47}, 2369 (1983)
[Bull. Acad. Sci. USSR, Phys. Ser. {\bf 47}, 81 (1983)].
\bibitem{Ganichev06} S.D.~Ganichev, V.V.~Bel'kov, S.A.~Tarasenko,
S.N.~Danilov, S.~Giglberger, Ch.~Hoffmann, E.L.~Ivchenko,
D.~Weiss, W.~Wegscheider, C.~Gerl, D.~Schuh, J.~Stahl,
J.~De~Boeck, G.~Borghs, and W.~Prettl, {\it Zero-bias spin
separation}, Nature Phys. {\bf 2}, 609 (2006).
\end{thebibliography}
\end{document}